\documentclass{article}
\usepackage{graphicx} 
\usepackage{amsmath}
\usepackage{amssymb}
\usepackage{cite}
\usepackage{geometry}
\title{Holographic dark energy with Granda-Oliveros cutoff and ansatz based approach : An extended look}
\author{%
    Oem Trivedi $^{1}$\thanks{oem.t@ahduni.edu.in} and
    Maxim Khlopov$^{2,3,4}$\thanks{khlopov@apc.in2p3.fr}, 
}

\date{%
    \small
    $^{1}$International Centre for Space Sciences and Cosmology, Ahmedabad University, Ahmedabad 380009, India\\
    $^{2}$Research Institute of Physics, Southern Federal University, 344090 Rostov-on-Don, Russia\\
    $^{3}$Virtual Institute of Astroparticle Physics, 75018 Paris, France\\
    $^{4}$Center for Cosmoparticle Physics Cosmion, National Research Nuclear University “MEPHI”, 115409 Moscow, Russia\\
    \today 
}

\begin{document}

\maketitle

\begin{abstract}
 Holographic dark energy models have proven to be a very interesting way to study various aspects of late-time acceleration of the universe. In this work we extensively study HDE models with the Granda-Oliveros cutoff with an ansatz based approach. We consider the Tsallis, Barrow and PLEC HDE models in this regard and consdier simple power law, emergent universe, intermediate and logamediate forms of  for the universe. Studying various cosmologically interesting parameters alongside the thermodynamical aspects in these models, we show that the Logamediate models are the best fit out of the other possibilites, followed by the emergent universe model, intermediate model and the simple power law models at the very last in terms of feasibility.     
\end{abstract}

\section{Introduction}
The discovery of the Universe's late-time acceleration marked a significant breakthrough in cosmology \cite{SupernovaSearchTeam:1998fmf}. Since then, extensive research has been devoted to understanding this expansion phenomenon. Various approaches have been explored, encompassing conventional methods like the Cosmological constant \cite{Weinberg:1988cp,Lombriser:2019jia,Padmanabhan:2002ji}, unconventional theories like Modified gravity \cite{Capozziello:2011et,Nojiri:2010wj,Nojiri:2017ncd}, and scenarios involving scalar fields driving late-time cosmic acceleration \cite{Zlatev:1998tr,Tsujikawa:2013fta,Faraoni:2000wk,Gasperini:2001pc,Capozziello:2003tk,Capozziello:2002rd,Odintsov:2023weg}. Additionally, quantum gravity theories such as Braneworld cosmology in string theory, loop quantum cosmology, and asymptotically safe cosmology have contributed to addressing the cosmic-acceleration puzzle \cite{Sahni:2002dx,Sami:2004xk,Tretyakov:2005en,Chen:2008ca,Fu:2008gh,Bonanno:2001hi,Bonanno:2001xi,Bentivegna:2003rr,Reuter:2005kb,Bonanno:2007wg,Weinberg:2009wa}. However, discrepancies persist, exemplified by the famous Hubble tension, reflecting the limitations of our current understanding of the universe, as evidenced by conflicting measurements of the Hubble constant \cite{Planck:2018vyg,riess2019large,riess2021comprehensive}. Thus, the present epoch of the universe poses profound questions, promising advancements in gravitational physics to deepen our comprehension of cosmology.

Among the various proposed solutions for the issues of late time acceleration is the idea of appplying the notions of the holographic principle \cite{tHooft:1993dmi,Susskind:1994vu}to cosmology. This principle posits that a system's entropy is determined not by its volume but by its surface area \cite{Bousso:1999xy}. Such an exotic form of DE also piques interest given the recent results from DESI \cite{DESI:2024lzq,DESI:2024mwx,DESI:2024uvr} which suggest a departure from $\Lambda$CDM cannot be completely ruled out. The initial work on HDEs carried out in \cite{Cohen:1998zx} proposed through a QFT approach that a short-distance cutoff is linked to a long-distance cutoff due to black hole formation limitations. Specifically, if $\rho$ denotes the quantum zero-point energy density from a short-distance cutoff, the total energy within a region of size $L$ should not exceed the mass of a black hole of the same size, yielding the inequality $L^3\rho\leq LM_{\text{pl}}^2$. The maximum allowable value for the infrared cutoff ($L_{_{\text{IR}}}$) precisely satisfies this inequality, leading to the relationship:

\begin{equation}\label{simphde}
\rho=3 c^2L_{_{\text{IR}}}^{-2},
\end{equation}

where $c$ is an arbitrary parameter, and we are using $m_{p} = 1 $ units.

This principle has found widespread application in cosmology, particularly in elucidating the late-time dark energy era, termed holographic dark energy (HDE) (for a comprehensive review, see \cite{Wang:2016och}). In this framework, the infrared cutoff, $L_{_{\text{IR}}}$, is cosmologically grounded and serves as the IR cutoff for a specific HDE model. Numerous studies have explored holographic dark energy from diverse perspectives in recent years \cite{Nojiri:2017opc,Oliveros:2022biu,Granda:2008dk,Khurshudyan:2016gmb,Wang:2016och,
Khurshudyan:2016uql,Belkacemi:2011zk,
Zhang:2011zze,Setare:2010zy,Nozari:2009zk,Sheykhi:2009dz,
Xu:2009xi,Wei:2009au,Setare:2008hm,Saridakis:2007wx,Setare:2006yj,
Felegary:2016znh,Dheepika:2021fqv,Nojiri:2005pu,Nojiri:2021iko,Nojiri:2020wmh,Trivedi:2024dju,Trivedi:2024rhp}. Over recent decades, several alternative forms of HDE have been proposed. For instance, Tsallis HDE models incorporate Tsallis' corrections to the standard Boltzmann-Gibbs entropy, resulting in the equation \begin{equation} \label{rtsa}
    \rho_{\Lambda} = 3 c^2 L^{-(4 - 2\sigma)}
\end{equation}, where $\sigma$ is the Tsallis parameter, assumed positive \cite{Tavayef:2018xwx}, with the simple HDE recovered in the limit $\sigma \to 1$. Conversely, Barrow's modification of the Bekenstein-Hawking formula

 led to Barrow HDE models described by the energy density \begin{equation} \label{rbar}
    \rho_{\Lambda} = 3 c^2 L^{\Delta - 2}
\end{equation}, where $\Delta$ is the deformation parameter \cite{Saridakis:2020zol}, capped at $\Delta = 1 $, and the simple HDE is regained in the limit of $\Delta \to 0$. Another very promising idea which has its origins in high energy physics is the generalized uncertainty principle, and taking inspiration from those theories, one can obtain Power law corrections to the entanglement entropy of black holes which result in another HDE form given by \begin{equation} \label{4}
    \rho_{\Lambda} = 3 L^{-2} m_{p}^2 \left( c^2 - \frac{\delta}{3} L^{2 - \gamma} \right)
\end{equation}, with $\gamma$ and $\delta$ representing parameters associated with underlying corrections \cite{Das:2007mj}. Such models are called as power law entropy corrected HDE models or PLECHDE. One sees that in the limit of $\delta \to 0 $, one recovers the simple HDE model while $\delta $ is usually constrained to be as $ 0 < \delta < 4$ \footnote{One can also make the case that in the limit of $\gamma \to \infty $, one also recovers the simple HDE but this is not a physical limit which can be reached in such models. Indeed it has been successfully argued that there is a limit to the values that $\gamma$ can take, which has been found to be $2 < \gamma < 4$ \cite{Das:2007mj,Radicella:2010ss}}. In this work we would like to discuss the various forms of cosmological evolution in the Tsallis, Barrow and PLEC HDE models with an ansatz based approach. In particular we would like to discuss the feasibility of a simple power law, intermediate law, emergent universe and Logamediate law evolutions. In Section II we shall discuss the basic formulations in this regard, while in section III, in its various subsections, we discuss the feasibility of the all the evolution schemes considered. We finally conclude our work in section IV. 
\section{Basic formulations }
We would like to consider an interacting dark sector in our work here. One can explore an interaction between dark energy and dark matter by incorporating the following continuity equations:

\begin{equation} \label{contd}
\dot{\rho}_{\Lambda} + 3 H \rho_{\Lambda} (1+ w_{\Lambda}) = - Q 
\end{equation}

\begin{equation} \label{contm}
\dot{\rho}_{m} + 3 H \rho_{m} (1+ w_{m}) = Q
\end{equation}

Here, $\rho_{\Lambda}$ and $\rho_m$ represent the energy densities of dark energy and dark matter, respectively. The parameter $w_{\Lambda}$ denotes the equation of state for dark energy, while $w_m$ represents the equation of state for dark matter. The term $Q$ describes the interaction between the two components.

For the sake of simplicity, we focus solely on the contributions of dark energy and dark matter to cosmology. The interaction term $Q$ can take linear or nonlinear forms. Some examples include:

\begin{equation} \label{q1}
Q = 3Hb (\rho_{\Lambda} + \rho_{m} )
\end{equation}

\begin{equation} \label{[q2]}
Q = 3Hb \frac{\rho_{\Lambda}^2}{\rho_{\Lambda} \rho_{m}}
\end{equation}

In these expressions, $b$ is a coefficient determining the strength of the interaction. These linear and nonlinear forms of the interaction term offer different possibilities to explore the dynamics of the interacting dark sectors. Before considering the ansatz now, it is first important that we do some analysis which is independent of the ansatz and is general. We also define the fractional energy density for DE and DM as \begin{equation} \label{omegad} 
    \Omega_{\Lambda} = \frac{\rho_{\Lambda}}{3 H^2}
\end{equation}
\begin{equation} \label{omegam}
    \Omega_{m} = \frac{\rho_{m}}{3 H^2}
\end{equation}
where $\rho_{\Lambda}$ would depend upon the model of HDE chosen by us. Now the cutoff we know very well is given by \begin{equation} \label{gocutoff}
L = (\alpha H^2 + \beta \dot{H})^{-1/2}
\end{equation}
It is also interesting here to note that in order to have a time varying dark energy, one does not always need to rely only on exotic models like HDEs, Scalar field paradigms etc. A fascinating possibility in this regard is the considertion of unstable dark matter alongside a cosmological constant form of dark energy \cite{Doroshkevich:1984gw,doroshkevich1984formation,doroshkevich1985fluctuations,doroshkevich1988cosmological,doroshkevich1989large}. Even such a form of the dark sector can produce the effect of a time varying dark energy scenario and in future works, it may be of interest to consider HDEs alongside unstable DM as well. 
\\
\\
We start off with the simple HDE model, \eqref{simphde} and take the derivative of the energy density to have \begin{equation} \label{r1}
    \dot{\rho}_{\Lambda} = -2 \rho_{\Lambda} \frac{\dot{L}}{L} 
\end{equation}
Here we would now need an expression for $-\frac{\dot{L}}{L}$ to proceed further, for which we use \eqref{gocutoff} to write \begin{equation} \label{leq}
    -\frac{\dot{L}}{L} = \frac{1}{2} L^2 (2 \alpha H \dot{H} + \beta \ddot{H} ) 
\end{equation}
Now this is the important part, as from here we can relate the dimensionless dark energy paramter with the help of \eqref{omegad} to have \begin{equation}
    L^2 = \frac{c^2}{\Omega_{\Lambda} H^2} 
\end{equation}
Using this we can write \begin{equation} \label{r11}
    \dot{\rho_{\Lambda}} = \frac{\rho_{\Lambda} c^2}{H^2 \Omega_{\Lambda}} (2 \alpha H \dot{H} + \beta \ddot{H} )
\end{equation}
This equation takes into account the fractional energy density parameter and relate it with the time derivative of the energy density. From here we can now use the continuity equation \eqref{contd} to have \begin{equation} \label{w1}
    w_{\Lambda} = -1 - \frac{1}{ 3 H \Omega_{\Lambda} } \Bigg[ \frac{c^2 (2 \alpha H \dot{H} + \beta \ddot{H} }{H^2} + \frac{Q}{3 H^2} \Bigg]
\end{equation}
From the EOS parameter \eqref{w1}, we can get the squared sound speed as \begin{equation} \label{vs1}
    v_{s}^2 = \frac{\dot{p_{\Lambda}} }{\dot{\rho_{\Lambda} }} = w_{\Lambda} + \frac{\dot{w_{\Lambda}} \rho_{\Lambda} }{\dot{\rho_{\Lambda} }}
\end{equation} 
We can now similarly write these quantities for other HDE models as well. Let's consider firstly the Tsallis model with \eqref{rtsa}, with its derivative being \begin{equation} \label{r2}
   \dot{\rho_{\Lambda}} =  - (4-2\sigma) \rho_{\Lambda} \frac{\dot{L}}{L}  
\end{equation} 
We can now define the fractional energy density parameter for the Tsallis model with \eqref{omegad}, from which we can write \begin{equation} \label{l2t}
    L^2 = \left( \frac{c^2}{\Omega_{\Lambda} H^2 } \right)^{\frac{1}{2 -\sigma}}
\end{equation}
Using this in \eqref{leq}, we can write \eqref{r2} as \begin{equation} \label{r21}
    \dot{\rho_{\Lambda}} = \frac{(4- 2 \sigma) \rho }{2} \left( \frac{c^2}{\Omega_{\Lambda} H^2 } \right)^{\frac{1}{2 -\sigma}} ( 2 \alpha H \dot{H} + \beta \ddot{H} )  
\end{equation}
We can then write using \eqref{contd}, \begin{equation} \label{wtsa}
    w_{\Lambda} = -1 - \frac{1}{3 H \Omega_{\Lambda}} \Bigg[ \frac{(4- 2 \sigma) \Omega_{\Lambda} }{2} \left( \frac{c^2}{\Omega_{\Lambda} H^2 } \right)^{\frac{1}{2 -\sigma}} ( 2 \alpha H \dot{H} + \beta \ddot{H} ) + \frac{Q}{3 H^2}  \Bigg]
\end{equation}
We note that in the limit $\sigma \to 1$, \eqref{wtsa} reduces to \eqref{w1}, as would be expected as in the limit of $\sigma \to 1$, Tsallis thermodynamics reduces to the usual thermodynamics. In a similar way, we can derive for the Barrow model \eqref{rbar}, to get the evolution of the energy density and the EOS parameter of dark energy as \begin{equation} \label{r3}
    \dot{\rho_{\Lambda}} = \frac{(2- \Delta) \rho }{2} \left( \frac{c^2}{\Omega_{\Lambda} H^2 } \right)^{\frac{2}{2 -\Delta}} ( 2 \alpha H \dot{H} + \beta \ddot{H} )
\end{equation}
\begin{equation} \label{wbar}
     w_{\Lambda} = -1 - \frac{1}{3 H \Omega_{\Lambda}} \Bigg[ \frac{(2- \Delta) \Omega_{\Lambda} }{2} \left( \frac{c^2}{\Omega_{\Lambda} H^2 } \right)^{\frac{2}{2 -\Delta}} ( 2 \alpha H \dot{H} + \beta \ddot{H} ) + \frac{Q}{3 H^2}  \Bigg]
\end{equation}
Where we note that in the limit of the deformation parameter $\Delta \to 0$, we recover the formulations for the simple HDE as one would expect. Finally, for the PLECHDE model we get \begin{equation} \label{r4}
    \dot{\rho_{\Lambda}} = \left( \frac{(2 \rho - (\gamma -2 ) \delta L^{-\gamma } ) }{2} \right) \left( \frac{3 c^2 - \delta L^{2 - \gamma }}{3 H^2 \Omega_{\Lambda}} \right) ( 2 \alpha H \dot{H} + \beta \ddot{H} ) 
\end{equation}
\begin{equation} \label{wplec}
    w_{\Lambda} = -1 - \frac{1}{3 H \Omega_{\Lambda}} \Bigg[ \left( \frac{(2 \Omega_{\Lambda} - (\gamma -2 ) \delta L^{-\gamma } ) }{2} \right) \left( \frac{3 c^2 - \delta L^{2 - \gamma }}{3 H^2 \Omega_{\Lambda}} \right) ( 2 \alpha H \dot{H} + \beta \ddot{H} ) + \frac{Q}{3 H^2}  \Bigg]
\end{equation}
Again we see that in the limit of $\delta \to 0$, we recover the formulations of the simple HDE. 
\\
\\
It is also important to discuss another point which could help ensure consistency of holographic dark energy models and that is thermodynamics. The thermodynamical implications and in particular, examining the validity of holographic dark energy models in the face of the generalized second law of thermodynamics was recently explored in \cite{Mamon:2020wnh}. It is well known that thermodynamical analysis of the gravity theory is an exciting research
topic in the cosmological context and the thermodynamical properties which hold for a black hole are equally valid for a cosmological horizon. In addition, the first
law of thermodynamics which holds in a black hole horizon can also be derived from the first Friedmann equation in the FLRW universe when the universe is bounded by an apparent horizon. This provides well motivation to select the apparent
horizon as the cosmological horizon in order to examine the thermodynamic properties of any cosmological model. Motivated by the above arguments, one can consider the universe as a thermodynamic system that is bounded by the
cosmological apparent horizon with the radius \cite{Bak:1999hd} \begin{equation} 
    r_{h} = \left( H^2 + \frac{k}{a^2} \right)^{-1/2}
\end{equation}
where for the case of the flat universe $(k=0)$, one can then write \begin{equation} \label{rh}
    r_{h} = \frac{1}{H}
\end{equation}
Considering that the major contributions to the entropy of the universe is from the dark sectors besides the cosmological horizon, we can then write that \begin{equation}
    S_{tot} = S_{\Lambda} + S_{m} + S_{h}
\end{equation}
where $S_{h} $ is the horizon entropy.  The time evolution of $S_{\Lambda} $ and $S_{m} $ can be written using the first law of thermodynamics as \begin{equation} \label{firstlawd}
    T \dot{S}_{\Lambda} = \dot{E}_{\Lambda} + p_{\Lambda} \dot{V}
\end{equation}
\begin{equation} \label{firstlawm}
    T \dot{S}_{m} = \dot{E}_{m} + p_{m} \dot{V}
\end{equation}
Where $V = \frac{4 \pi r_{h}^3 }{3}$ is the horizon volume, with $$E_{\Lambda}= \frac{4 \pi r_{h}^3 \rho_{\Lambda} }{3}$$ and $$E_{m}= \frac{4 \pi r_{h}^3 \rho_{m} }{3}$$
with the time evolutions of the energy densities being given by the continuity equations \eqref{contd} and \eqref{contm}. The temperature T of the apparent horizon is given by \begin{equation}
    T = \frac{1}{2 \pi r_{h}} = \frac{H}{2 \pi}
\end{equation}  
The entropy of the horizon would depend on what model of HDE we are considering. In the case of the simple HDE, we have the usual Benkenstein hawking entropy at work, which we can write as \begin{equation} 
    S_{h} = \frac{A}{A_{0}}  
\end{equation} 
where $A_{0} = \frac{1}{4 G}$ is the Planck area and we can write the above equation as \begin{equation} \label{sbenk}
    S_{h} = \gamma r_{h}^2
\end{equation}
 where $\gamma$ here is a positive constant ( in this case equal to $ \pi / G $ but the utility of such an expression for $S_{h} $ will shortly be clear ). When we are talking of the Barrow entropy, we have \begin{equation}
     S_{h} = \left( \frac{A}{A_{0}} \right)^{1 + \Delta / 2}
 \end{equation}
 where again symbols have similar meanings as before and now we can express this instead in terms of the horizon radius as \begin{equation} \label{sbar}
     S_{h} = \gamma_{b} r_{h}^{2 + \Delta }
 \end{equation}
where $$\gamma_{b} = \left( \frac{4 \pi}{A_{0}} \right)^{1 + \Delta / 2 } $$ which will again just be a positive constant. For the Tsallis HDE scenario, the horizon entropy will be similarly given by \begin{equation} \label{stsa}
    S_{h} = \gamma r_{h}^{2 \sigma}
\end{equation}
where $\gamma_{t}$ is again a constant in terms of the Planck area and the Tsallis parameter. We can now write the time evolution of entropy for a universe with simple HDE using \eqref{firstlawd}-\eqref{firstlawm}  \begin{equation} \label{simpleentropy}
 \dot{S} = \frac{2 \pi}{H} \Bigg[ \frac{4 \pi}{3 H^3} ( 3 c^2 ( 2 \alpha H \dot{H} + \beta \ddot{H} ) + Q) - 12 \Omega_{m} \pi (1 + w_{m} ) \Bigg] - \frac{2 \gamma \dot{H} }{H^3}   
\end{equation}
In the case of the Tsallis scenario, we have the following \begin{equation} \label{tsallisentropy}
    \dot{S} = \frac{2 \pi}{H} \Bigg[ \frac{4 \pi}{3 H^3} \left( \frac{4 - 2 \sigma}{2} 3 H^2 \Omega_{\Lambda} \left( \frac{c^2}{H^2 \Omega_{\Lambda}} \right)^{\frac{1}{2 - \sigma}} (2 \alpha H \dot{H} + \beta \ddot{H} )  \right) + \frac{4 \pi Q}{3 H^3} - 12 \Omega_{m} \pi (1 + w_{m} )  \Bigg] - \frac{2 \gamma_{t} \sigma \dot{H}}{H^2 H^{2 \sigma - 1} }
\end{equation}
While in the case of Barrow HDE , we have \begin{equation} \label{barrowentropy}
    \dot{S} = \frac{2 \pi}{H} \Bigg[ \frac{4 \pi}{3 H^3} \left( \frac{2 - \Delta}{2} 3 H^2 \Omega_{\Lambda} \left( \frac{c^2}{H^2 \Omega_{\Lambda}} \right)^{\frac{2}{2 - \Delta}} (2 \alpha H \dot{H} + \beta \ddot{H} )  \right) + \frac{4 \pi Q}{3 H^3} - 12 \Omega_{m} \pi (1 + w_{m} )  \Bigg] - \frac{(2+\Delta) \gamma_{b}  \dot{H}}{H^2 H^{1 + \Delta} }
\end{equation}
\\
Again, we see that the formulas are consistent as \eqref{tsallisentropy} and \eqref{barrowentropy} reduce to \eqref{simpleentropy} in the case of $\sigma \to 1 $ and $\Delta \to 0$, respectively. With these formulations at hand, we can now proceed towards the main plots and results. 
\\
\\
\section{Different evolutions with G-O cutoff}
We have now developed enough machinery to tackle the models we want to. This time we consider not only the simple power law model \begin{equation} \label{simpowerans}
    a(t) = a_{0} t^n
\end{equation}
But also the emergent universe scenario with the ansatz \begin{equation} \label{emergentant}
    a(t) = a_{0} (\mu t + \lambda )^m
\end{equation}
And the logamediate scenario with the ansatz \begin{equation} \label{logamediateansatz}
    a(t) = a_{0} e^{A ((\log t)^{\alpha_{1}})} 
\end{equation}
where we denote the ansatz in \eqref{simpowerans}, \eqref{emergentant} and \eqref{logamediateansatz} as $a_{1}$, $a_{2}$ and $a_{3}$ respectively.
\\
\\
With this, we can start to consider the plots. We have set $c=1$ as that is fine for this cutoff and have considered a value of $\beta = 0.5$, but the value of $\alpha$ is not constrained and so we can think of ranges of $(\alpha,\beta)$ where the various holographic scenarios are viable or not considering $w$ and $v_{s}^2$. For the interaction terms, we keep the parameter $b=0.03$, We take $\Omega_{d} = 0.69$ and $\Omega_{m} = 0.258$, we take the various parameters in the ansatz fomrulaes as $n=3$, $m=2$, $\mu = 2$, $\lambda = 4$, $\alpha_{1}=4$, $A=0.02$ and $\gamma = 3$ ( the parameter in the PLEC model ). All these values are consistent with the theoretical needs of these models as seen in the literature.
\subsection{Simple power law evolution}
\begin{figure}[!ht]
    \centering
    \includegraphics[width=1\linewidth]{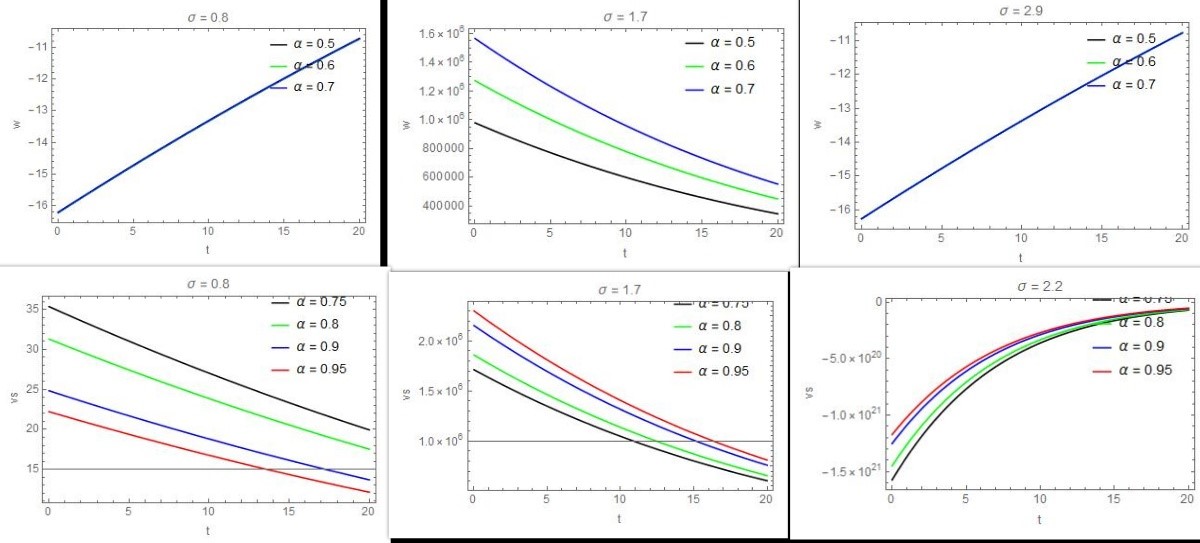}
    \caption{Plots of the Dark energy EOS and squared sound speed against time for linear interaction for simple power law in Tsallis HDE}
    \label{q1a1t}
\end{figure}

\begin{figure}[!ht]
    \centering
    \includegraphics[width=1\linewidth]{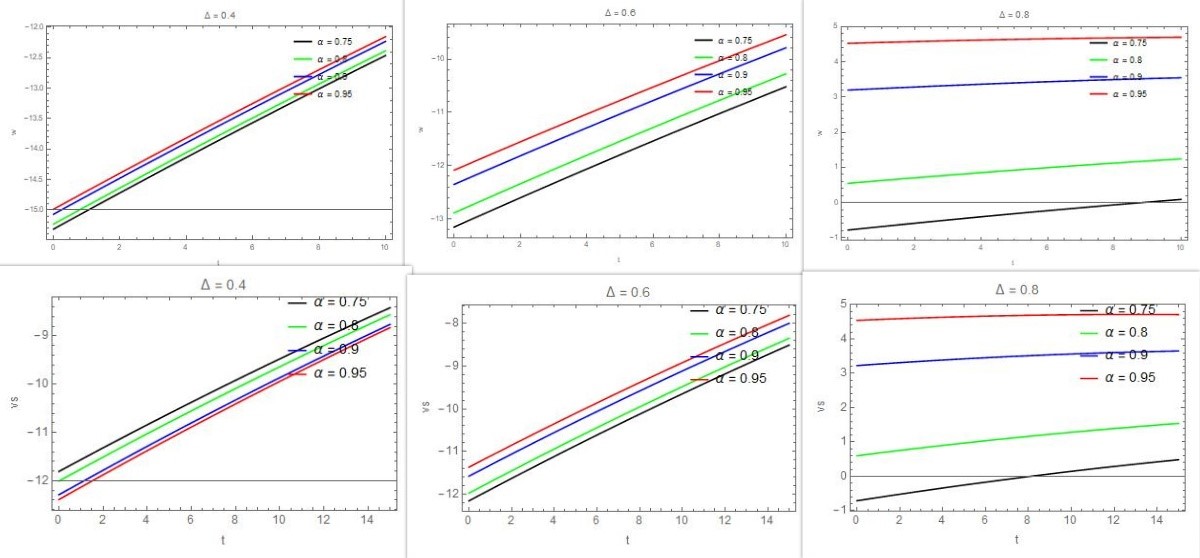}
    \caption{Plots of the Dark energy EOS and squared sound speed against time for linear interaction for simple power law in Barrow HDE}
    \label{q1a1b}
\end{figure}

\begin{figure}[!ht]
    \centering
    \includegraphics[width=1\linewidth]{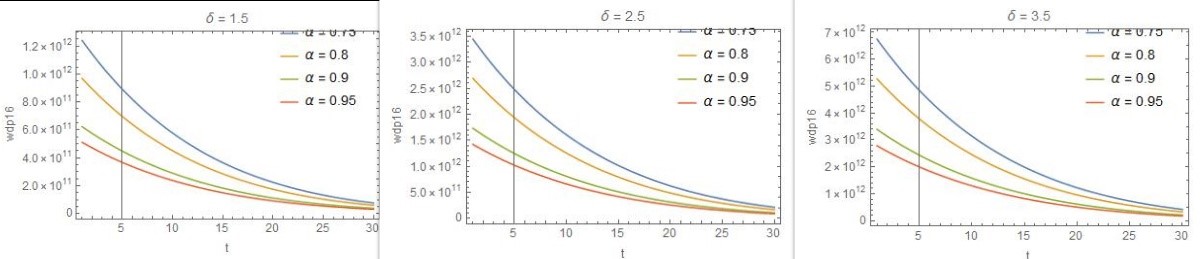}
    \caption{Plots of the Dark energy EOS and squared sound speed against time for linear interaction for simple power law in PLEC HDE}
    \label{q1a1p}
\end{figure}

\begin{figure}[!ht]
    \centering
    \includegraphics[width=1\linewidth]{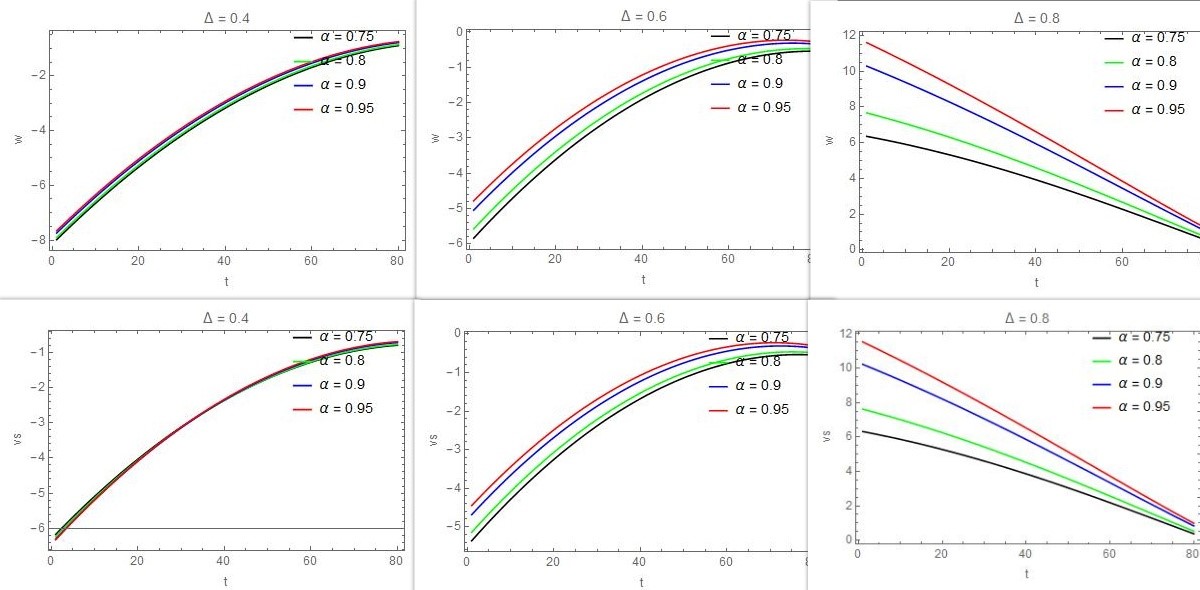}
    \caption{Plot of the dark energy EOS and squared sound speed against time for non-linear interaction with simple power law ansatz in Barrow HDE}
    \label{a1q2b}
\end{figure}

\begin{figure}[!ht]
    \centering
    \includegraphics[width=1\linewidth]{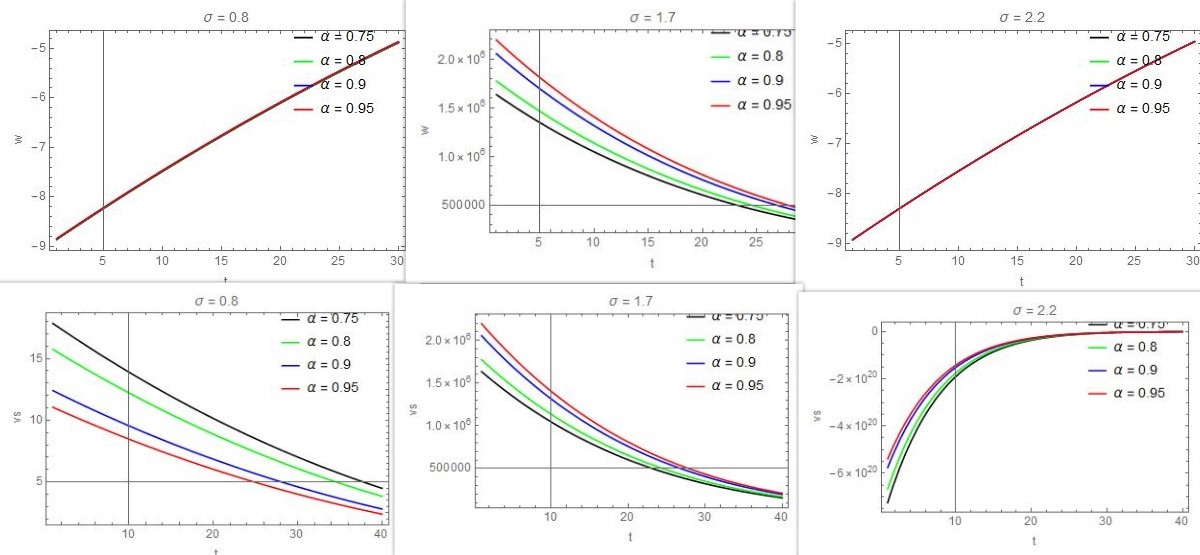}
    \caption{Plot of the dark energy EOS and squared sound speed against time for non-linear interaction with simple power law ansatz in Tsallis HDE}
    \label{a1q2t}
\end{figure}

\begin{figure}[!ht]
    \centering
    \includegraphics[width=1\linewidth]{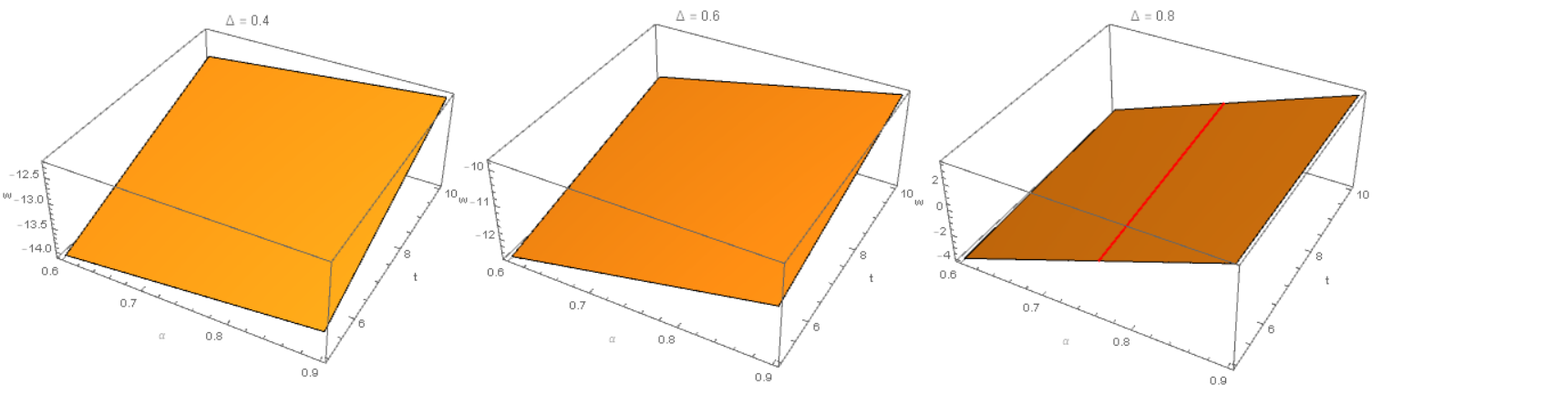}
    \caption{3d Plot of the dark energy EOS against time for linear interaction in the power law scenario for the Barrow model}
    \label{a1q13dwb}
\end{figure}

We start off with the Tsallis model, with the linear interaction \eqref{q1} and the ansatz \eqref{simpowerans}. The dark energy EOS and squared sound speed for this for three different values of the Tsallis parameter $\sigma$ are plotted agains time, where in each plot we have considered the range of $\alpha$ from $0.7 to 0.95$. The plots are displayed in figure \ref{q1a1t}.
We see that neither the squared sound speed nor the dark energy EOS are in any theoretical or observationally sound ranges, as the dark energy EOS takes absurdly negative or positive values while the squared sound speed is incredibly negative too. Hence, this scenario is not favoured much.
\\
\\
We then turn our attention to the Barrow model with the linear interaction \eqref{q1} and the power law ansatz \eqref{simpowerans}. Figure \ref{q1a1b} gives the plot of the DE EOS and squared sound speed for three different values of the deformation parameter $\Delta$, again in various ranges of $\alpha$. 
We see that neither the squared sound speed nor the dark energy EOS are in any theoretical or observationally sound ranges, as both the dark energy EOS and the squared sound speed both take highly negative values. Hence, this scenario is not favoured much. For the PLEC model with linear interaction and power law ansatz, we have plotted the EOS in figure \ref{q1a1p} and we see that the EOS is not viable in any case at all and so we have not went in for the sound speed in the case at all. Hence, here we can conclude that simple power law models with linear interaction are heavily disfavoured.
\\
\\
We now plot for the non-linear interaction scheme \eqref{[q2]}, where we firstly plot for the power law ansatz in the Barrow model. We see in figure \ref{a1q2b} that although in the linear scheme the Barrow model was completely dismissive of the simple power law model, in the non-linear scheme the dark energy EOS can be in viable ranges $\Delta = 0.4,0.6$. But the squared sound speed is only viable for $\Delta > 0.7$, which makes the model incompatible as a whole again but not as bad as with the linear scheme.  
In figure \ref{a1q2t} we have plotted for the simple power law and non linear interaction in the Tsallis model and we see that the EOS parameter is again not viable across various ranges of the free parameters and squared sound speed is only eventually becoming somewhat compatible for $\sigma > 2$. But as a whole, we see that this model is again not very favourable. For the PLEC model, the EOS parameter follows almost exactly the same plot and conclusions as in the linear interaction case shown in figure \ref{q1a1p} and so we have not plotted that again. So again we conclude that the simple power law models are quite disfavoured.
We also made a 3d plot for dark energy EOS in the Barrow model just for illustration purposes in figure \ref{a1q13dwb}.  
\subsection{Emergent universe evolution}
\begin{figure}[!ht]
    \centering
    \includegraphics[width=1\linewidth]{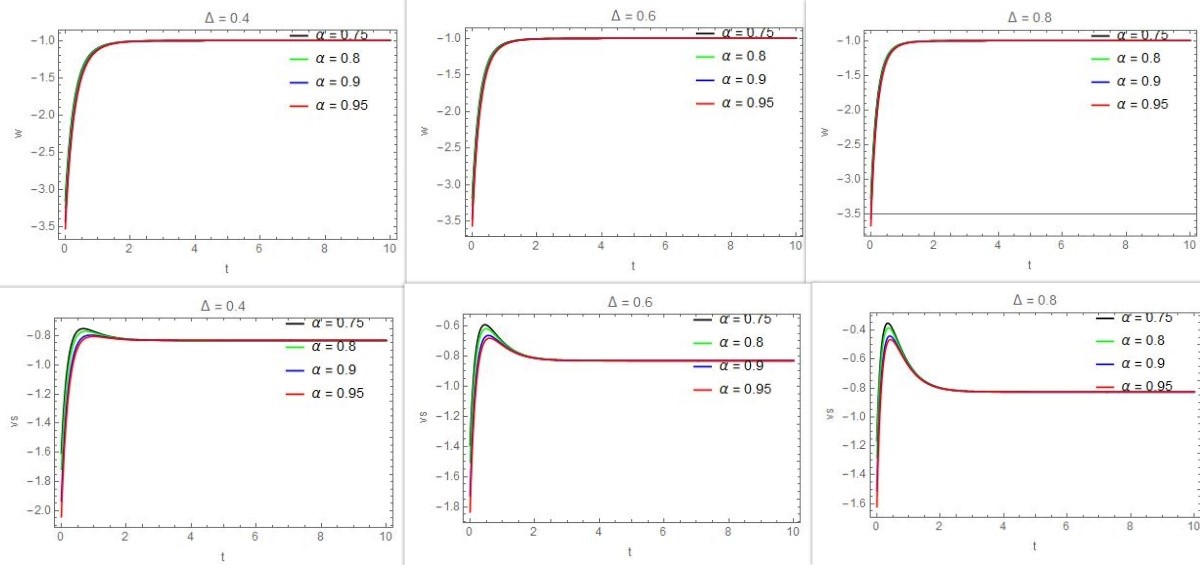}
    \caption{Plots of the Dark energy EOS and squared sound speed against time for linear interaction for emergent scenario ansatz in Barrow HDE}
    \label{q1a2b}
\end{figure}

\begin{figure}[!ht]
    \centering
    \includegraphics[width=1\linewidth]{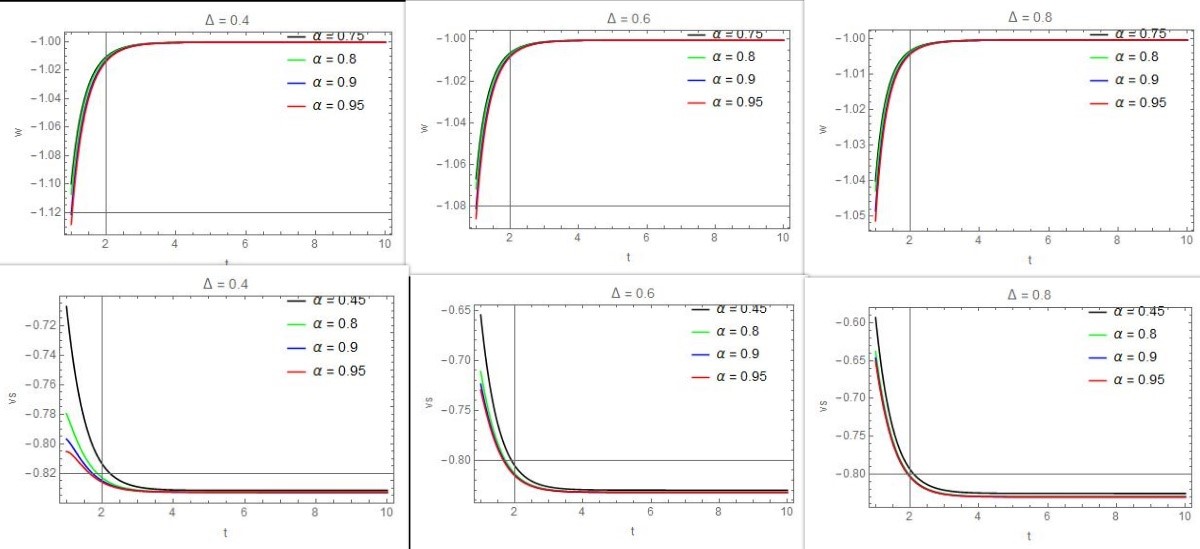}
    \caption{Plots of the Dark energy EOS and squared sound speed against time for non-linear interaction for emergent scenario ansatz in Barrow HDE}
    \label{q2a2b}
\end{figure}

\begin{figure}
    \centering
    \includegraphics[width=1\linewidth]{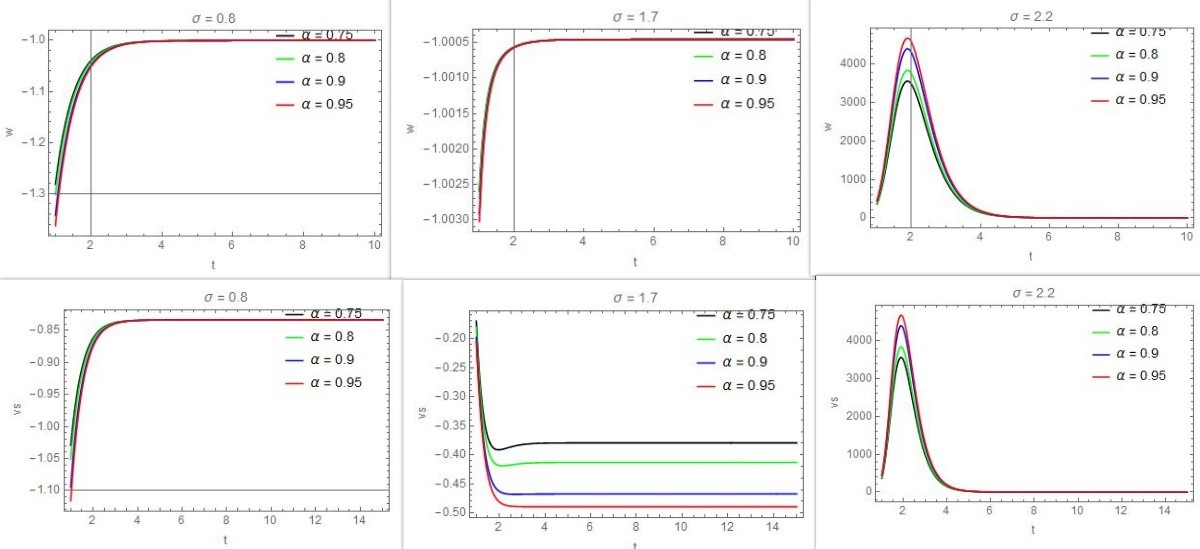}
    \caption{Plots of the Dark energy EOS and squared sound speed against time for non-linear interaction for emergent scenario ansatz in Tsallis HDE}
    \label{q2a2t}
\end{figure}

\begin{figure}[!ht]
    \centering
    \includegraphics[width=1\linewidth]{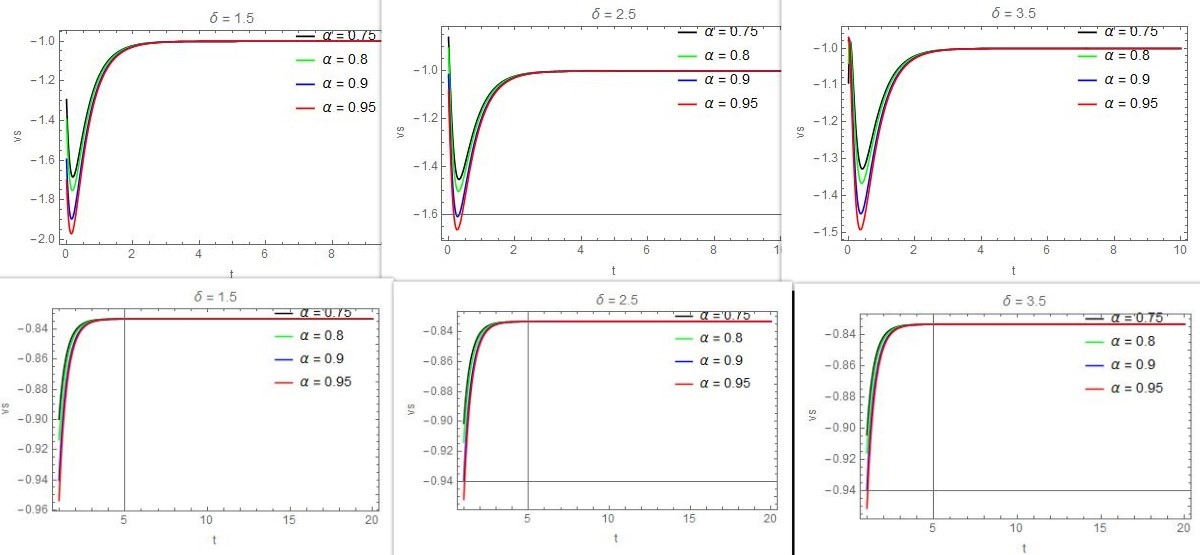}
    \caption{Plots of the Dark energy EOS and squared sound speed against time for non-linear interaction for emergent scenario ansatz in PLEC HDE}
    \label{q2a2p}
\end{figure}

\begin{figure}[!ht]
    \centering
    \includegraphics[width=1\linewidth]{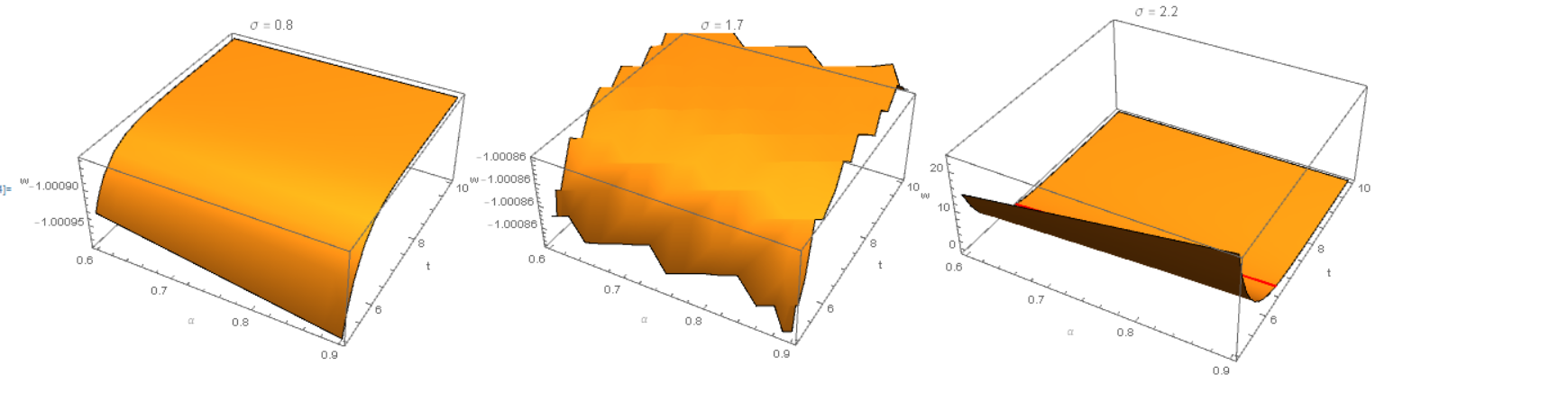}
    \caption{3d Plot of the dark energy EOS against time for linear interaction in the emergent scenario for the Tsallis model}
    \label{a2q23dwt}
\end{figure}

\begin{figure}[!ht]
    \centering
    \includegraphics[width=1\linewidth]{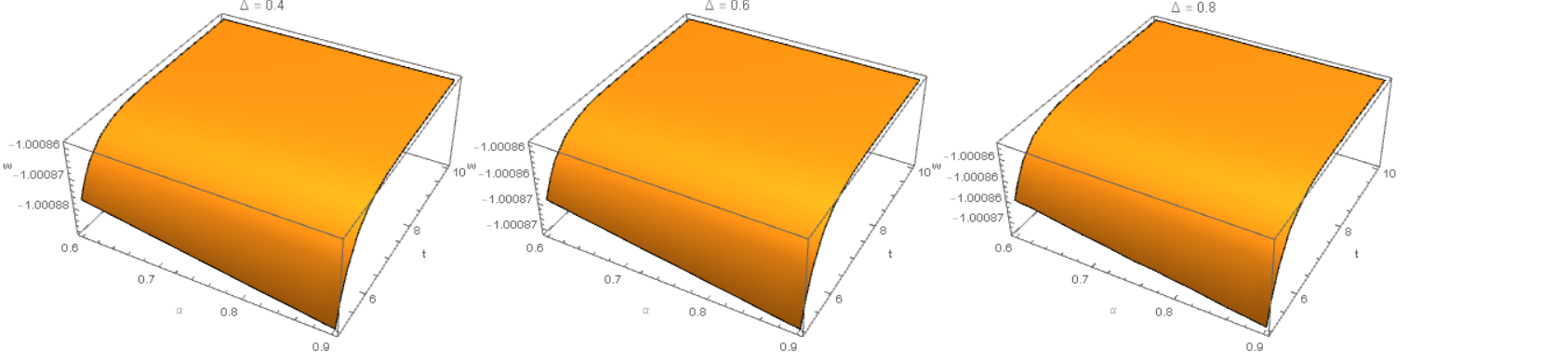}
    \caption{3d Plot of the dark energy EOS against time for linear interaction in the emergent scenario for the Barrow model}
    \label{a2q13dwb}
\end{figure}

We now consider the second ansatz, the emergent scenario \eqref{emergentant} with the linear interaction in the Barrow model and we have plotted the EOS and sound speed in figure \ref{q1a2b} for this. 
We see that in this scenario, the dark energy EOS is in the observationally viable range for various values of the deformation parameter and a wide range of values across $\alpha$ too but the squared sound speed is not viable in any of the range of $\alpha$ and $\Delta$.  
\begin{figure}[!ht]
    \centering
    \includegraphics[width=1\linewidth]{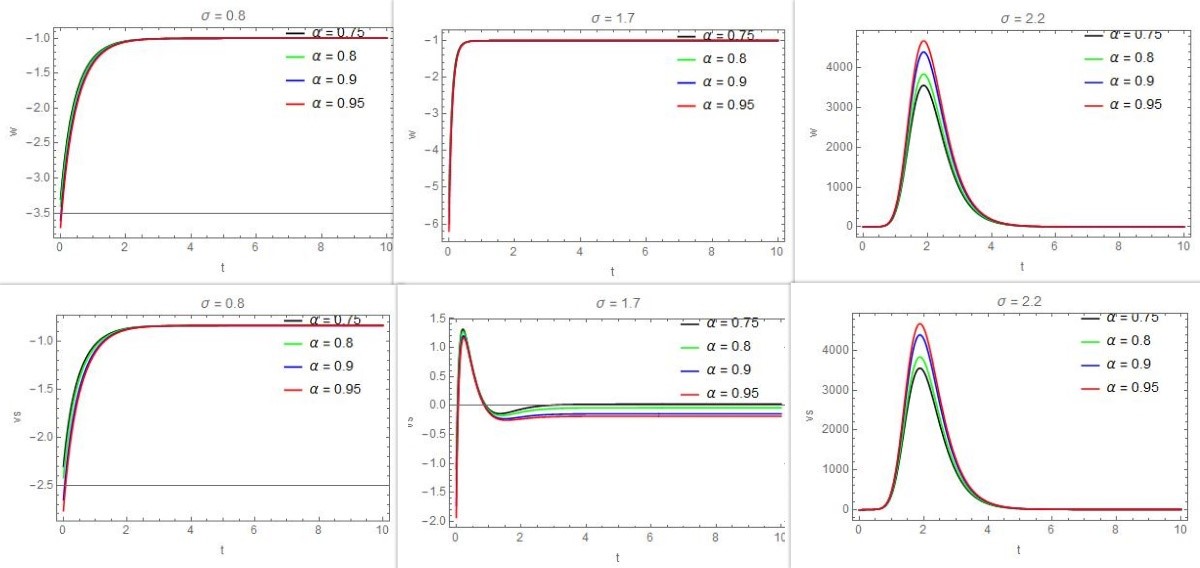}
    \caption{Plots of the Dark energy EOS and squared sound speed against time for linear interaction for emergent scenario ansatz in Tsallis HDE}
    \label{q1a2t}
\end{figure}
Plotting for the same ansatz in the linear interaction for the Tsallis model as in figure \ref{q1a2t}, we see that the EOS is viable for all range of $\alpha$ while for $\sigma > 2$, the squared sound speed is viable for $\alpha$ too and for $\sigma > 1.5$ the sound speed is still viable for $\alpha < 0.8$. Hence this model is quite viable. 
\begin{figure}[!ht]
    \centering
    \includegraphics[width=1\linewidth]{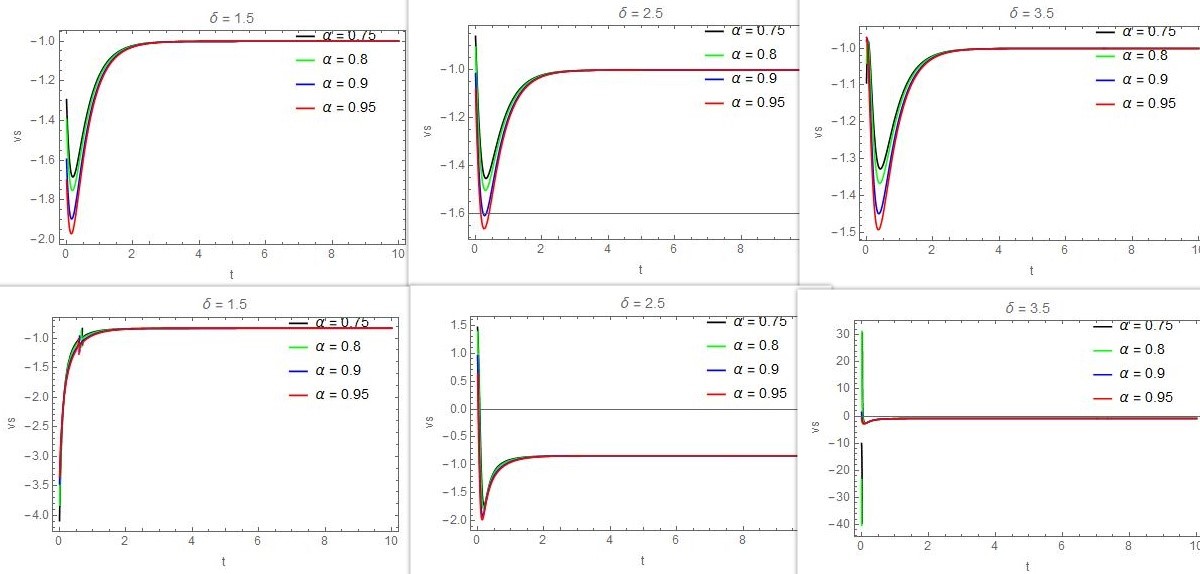}
    \caption{Plot of the dark energy EOS and squared sound speed against time for linear interaction with simple emergent scenario ansatz in PLEC HDE}
    \label{a2q1p}
\end{figure}
For the PLEC model with this ansatz in the linear interaction, plots have been made in figure \ref{a2q1p} and we see that the EOS is observationally viable in all parameter ranges, while the sound speed is viable for $\alpha > 0.9$ and $\delta >3$. So we can say that this is viable as well. Overall we can say that the emergent scenario models perform quite a lot better with linear interaction than the simple power law ones.
\\
\\
We have plotted for the non linear interaction with the emergent scenario ansatz for the Barrow model in figure \ref{q2a2b}. The plots show that while the EOS is viable for various ranges of $\Delta$ and $\alpha$, the squared sound speed is not. 
Plotting for the emergent scenario with non linear interaction in the Tsallis model in figure \ref{q2a2t}, we see that the EOS is viable for all ranges of $\alpha$ and $\sigma$ and the sound speed is also viable, but only for $\sigma > 2$. 
For the PLEC model with emergent ansatz and non linear interaction we see that while the EOS can be in viable ranges, the squared sound speed cannot as shown in figure \ref{q2a2p}. So as a whole while there are minor tweaks, the emergent universe models perform almost the same as in the non-linear interaction as they do in the linear interaction.   
We have also plotted the entropy for a viable value of the Tsallis parameter $\sigma > 2.2 $ ( where it showed both compatible w and $v_{s}^2$), in the emergent scenario. 
\begin{figure}[!ht]
    \centering
    \includegraphics[width=1\linewidth]{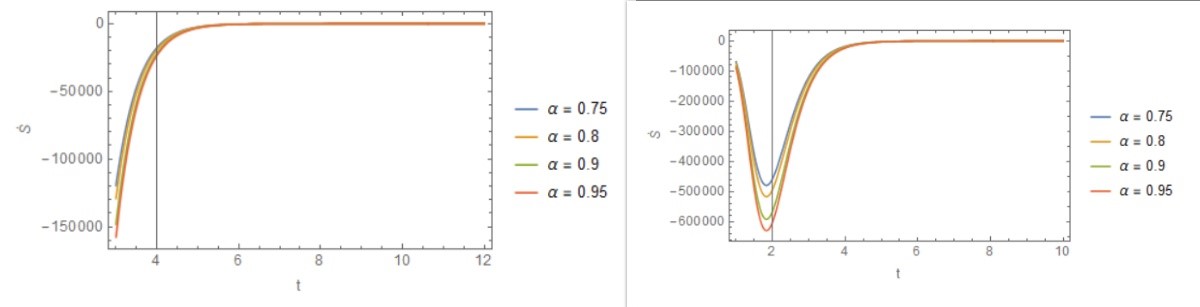}
    \caption{Plot of the time derivative of the universal entropy against time for viable value of the Tsallis parameter in emergent scenario}
    \label{sa2t}
\end{figure}
We have plotted this in figure \ref{sa2t} for both linear and non linear interactions, and we see that the generalized second law of thermodynamics can be satisfied as even though the plot starts off in the negative regime it eventually stays on for values $\sim 0$. So the emergent scenario with the Tsallis model for $\sigma > 2$ seems like a good model. Finally we have also made 3d plots of the dark energy EOS for the Barrow and Tsallis models in the emergent scenario in the linear interaction case in figures \ref{a2q23dwt} and \ref{a2q13dwb} just for illustrations. 
\subsection{Intermediate law evolution}
\begin{figure}[!ht]
    \centering
    \includegraphics[width=1\linewidth]{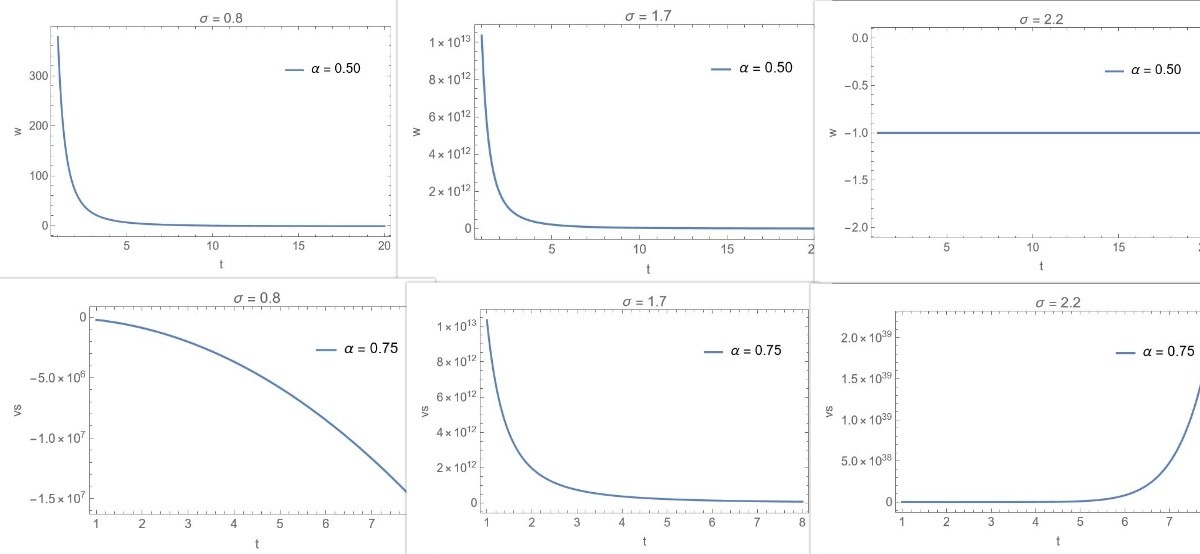}
    \caption{Plots of the EOS parameter of dark energy and squared sound speed against time for various values of $\sigma$ for the intermediate scenario with Tsallis HDE and linear interaction. The plot is the same for all viable values of $\alpha$}
    \label{a4q1t}
\end{figure}

\begin{figure}[!ht]
    \centering
    \includegraphics[width=1\linewidth]{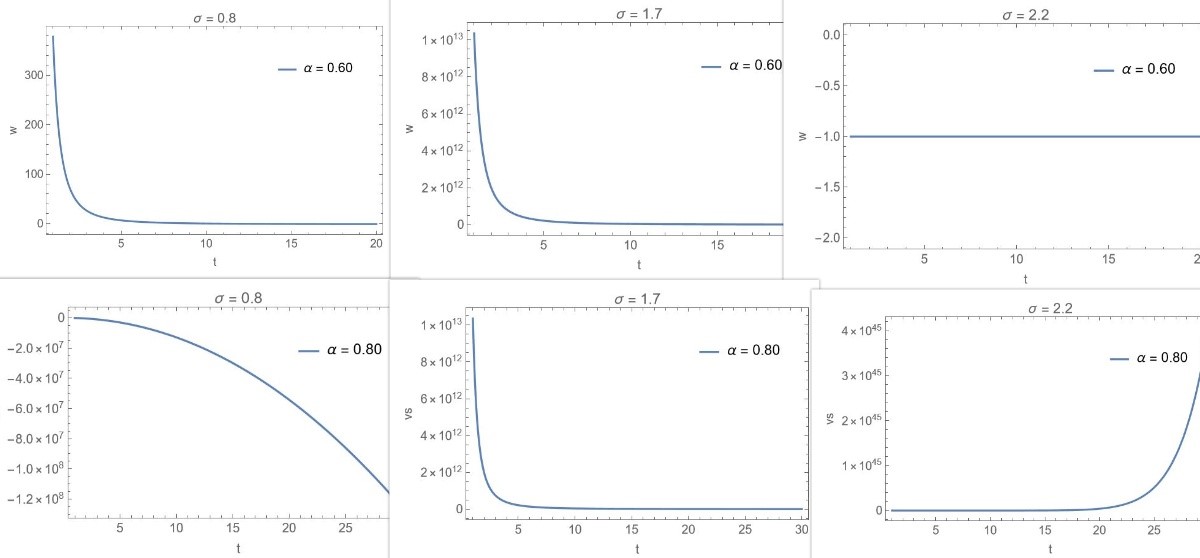}
    \caption{Plots of the EOS parameter of dark energy and squared sound speed against time for various values of $\sigma$ for the intermediate scenario with Tsallis HDE and non linear interaction.The plot is the same for all viable values of $\alpha$}
    \label{a4q2t}
\end{figure}

\begin{figure}[!ht]
    \centering
    \includegraphics[width=1\linewidth]{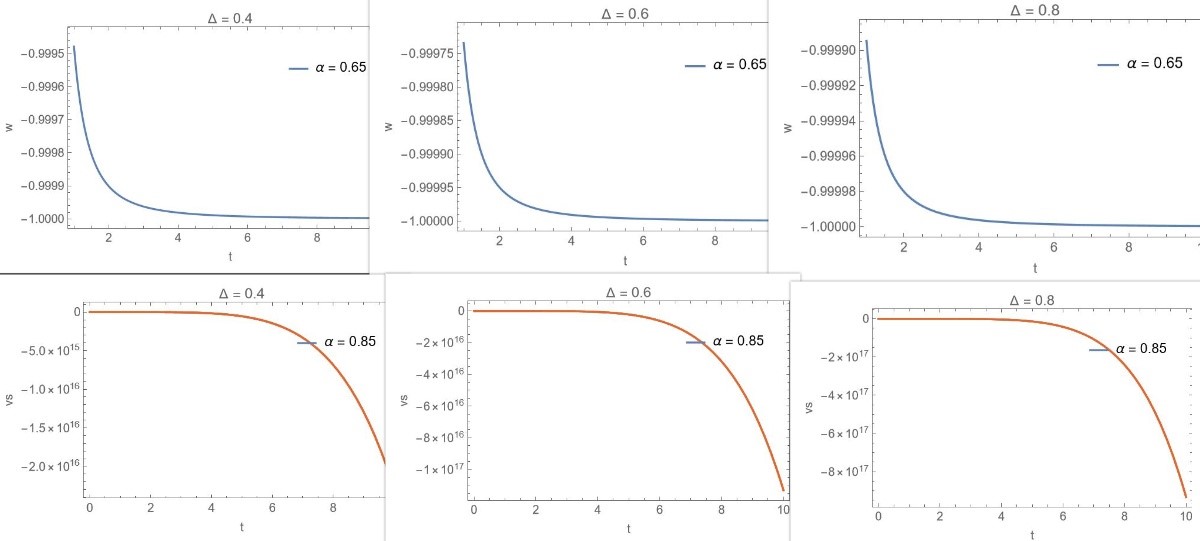}
    \caption{Plots of the EOS parameter of dark energy and squared sound speed against time for various values of $\Delta$ for the intermediate scenario with Barrow HDE and non linear interaction.The plot is the same for all viable values of $\alpha$ and also same for linear interaction too.}
    \label{a4q2b}
\end{figure}

\begin{figure}[!ht]
    \centering
    \includegraphics[width=1\linewidth]{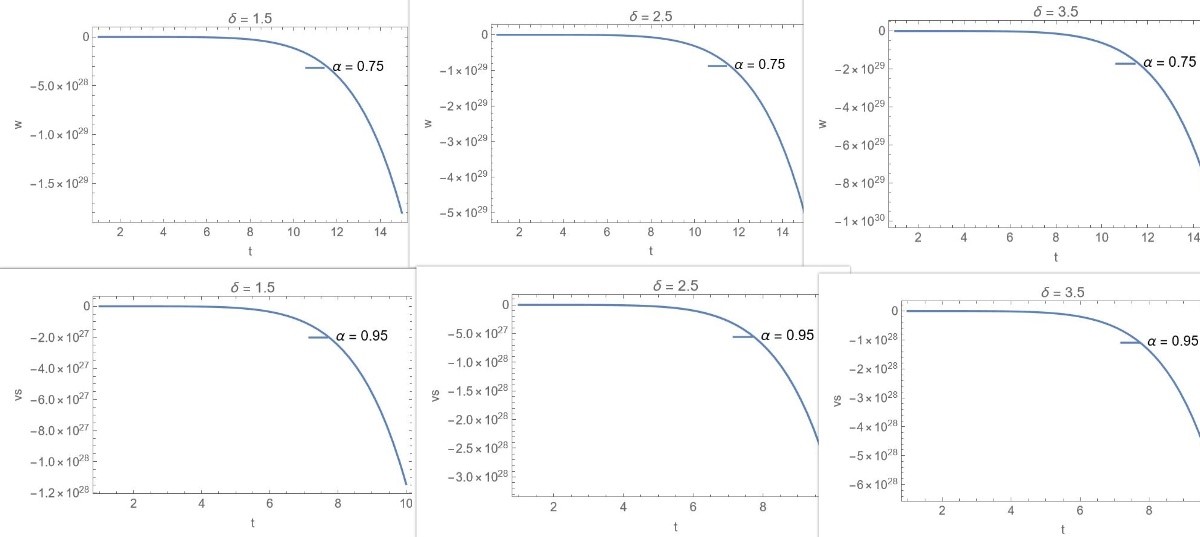}
    \caption{Plots of the EOS parameter of dark energy and squared sound speed against time for various values of $\delta$ for the intermediate scenario with PLEC HDE and linear interaction.The plot is the same for all viable values of $\alpha$ and also same for non linear interaction too.}
    \label{a4q1p}
\end{figure}

Plotting firstly for the Tsallis HDE scenario with linear interaction as in figure \ref{a4q1t}, we see that model gives viable values for the EOS parameter for $\sigma< 1$ and also for $\sigma > 2$. But the squared sound speed is not viable in any case as it is either too positive or too negative, across all ranges of the Tsallis parameter. So it seems that this scenario is not very favourable.
Plotting for the Tsallis HDE scenario with non linear interaction as in figure \ref{a4q2t}, we see that model gives viable values for the EOS parameter again for $\sigma< 1$ and also for $\sigma > 2$. But the squared sound speed is not viable once again in any case as it is either having very large positive or negative values for all ranges of the Tsallis parameter. So it seems that this scenario is not very favourable too. One thing we would notice in both these plots is that we have only plotted for one value of $\alpha$, the reason behind that is that interestingly the results are the same for all viable values of $\alpha$ (which corresponds to $\alpha \sim \mathcal{O}(1)$
Plotting for the Barrow model, we observed that the plots are quite the same for both linear and non linear interactions so we have just plotted for the non linear case in figure \ref{a4q2b}. We see that while the EOS parameter gives us viable values for all ranges of the deformation paraemter, the squared sound speed is incredibly negative for all ranges of $\Delta$ as well hereby making this scenario very unrealistic too. As the plots here for these parameters are the almost the same for the linear interaction too, we conclude similarly for that as well. 
Plotting for the PLEC HDE scenario we again noticed that both the linear and non linear cases of interaction were giving almost the same plots and so we have plotted here in figure \ref{a4q1p} for the linear interaction only. We see here that both the EOS parameter and the squared sound speed are incredibly unrealistic, for all ranges of the model parameter $\delta$ and so the scenario seems very unfavourable as well. Similar conclusions hold for the non linear interaction case here.
\subsection{Logamediate law evolution}
\begin{figure}[!ht]
    \centering
    \includegraphics[width=1\linewidth]{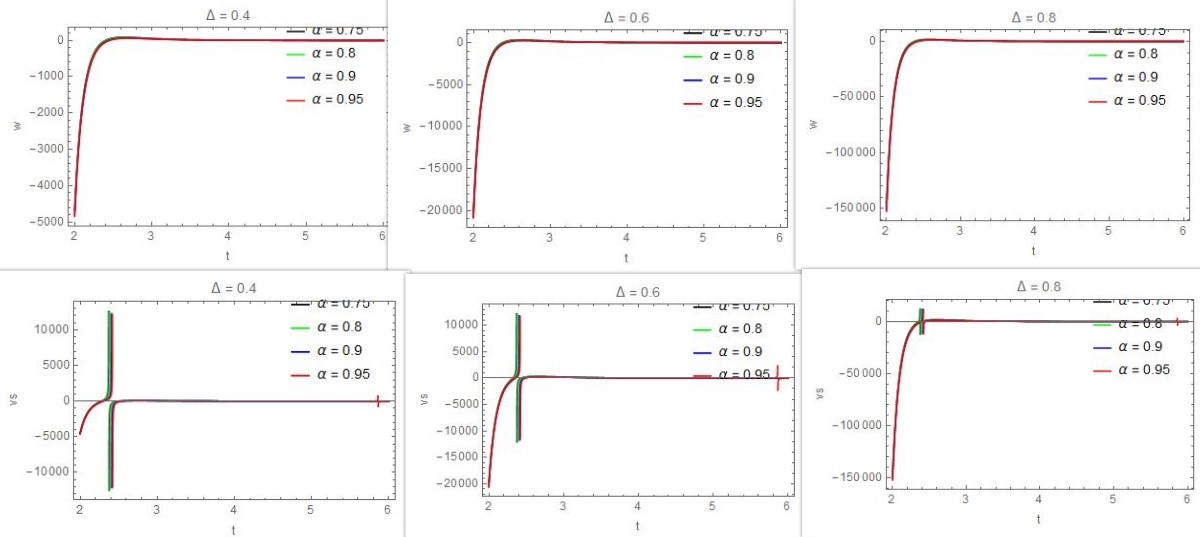}
    \caption{Plots of the Dark energy EOS and squared sound speed against time for linear interaction for Logamediate scenario ansatz in Barrow HDE}
    \label{q1a3b}
\end{figure}

\begin{figure}[!ht]
    \centering
    \includegraphics[width=1\linewidth]{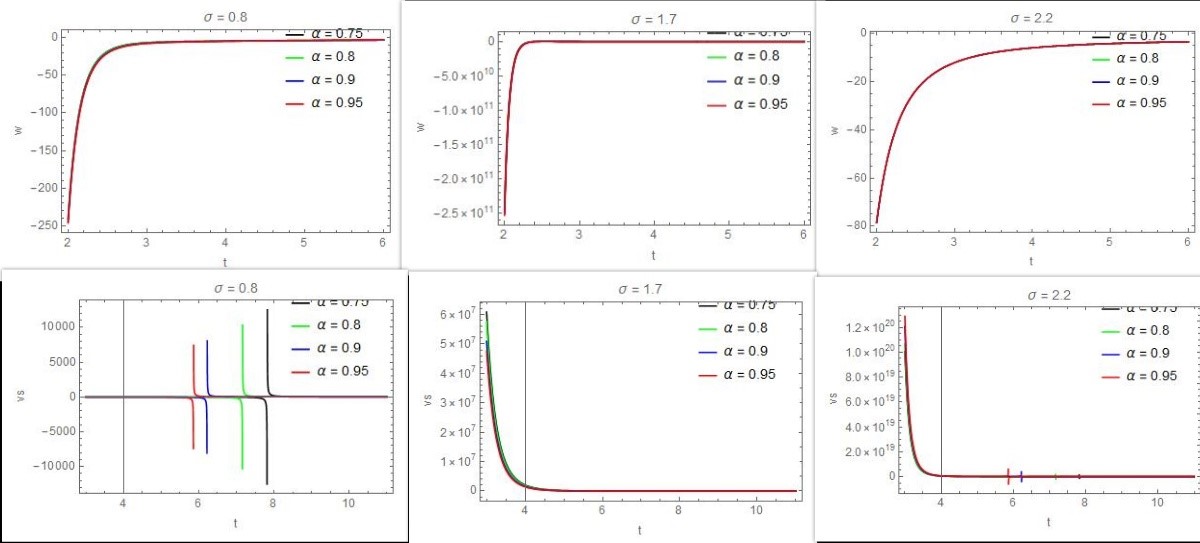}
    \caption{Plots of the Dark energy EOS and squared sound speed against time for linear interaction for Logamediate scenario ansatz in Tsallis HDE}
    \label{q1a3t}
\end{figure}

\begin{figure}[!ht]
    \centering
    \includegraphics[width=1\linewidth]{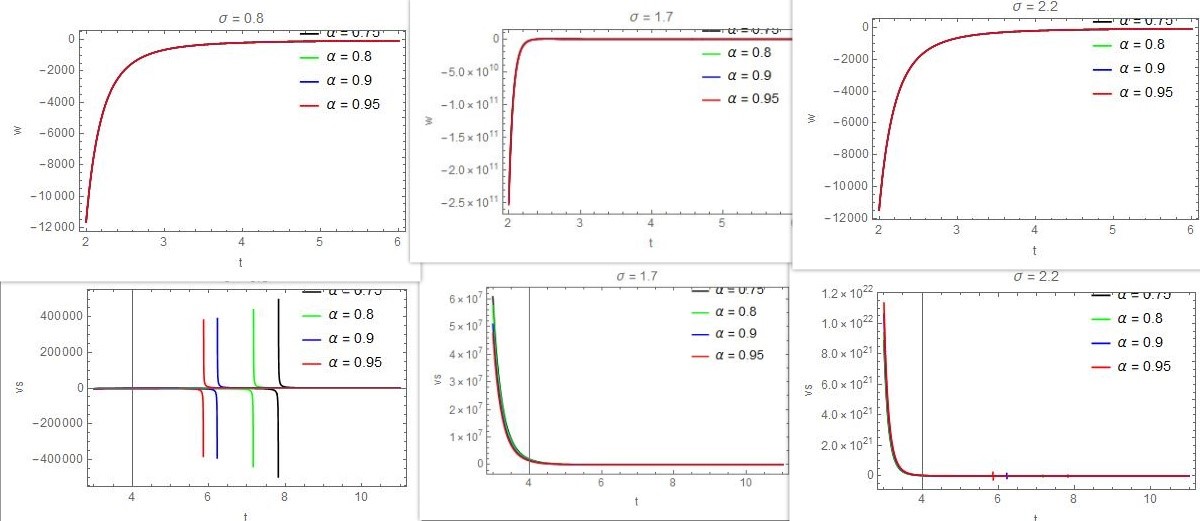}
    \caption{Plot of the dark energy EOS and squared sound speed against time for non-linear interaction with logamediate scenario ansatz in Tsallis HDE}
    \label{a3q2t}
\end{figure}

We have plotted similarly for Logamediate scenario \eqref{logamediateansatz} with linear interaction, where we plotted for the Barrow model as shown in figure \ref{q1a3b} and for the Tsallis model in figure \ref{q1a3t}.
What is interesting to note is that for both these models, the EOS of dark energy has the same behaviour, where it eventually hovers to value near -1 very quickly and then stays that way. Similarly, squared sound speed has similar behaviour too where we see that the sound speed takes a few jumps in both cases initially but then stays on to values $\sim 0$. Interestingly, the behaviour is exactly the same for the PLEC models and we have not plotted for that here because of that. So the logamediate ansatz with linear interaction produces viable models in all of Barrow, Tsallis and PLEC models, making it the most viable model we have seen so far. 
\\
\\
For the Logamediate case in the non-linear interactions, we find that the plots follow and are almost exactly the same as in the linear interaction case for all of Barrow, Tsallis and PLEC models. For example, we show the plot for the Tsallis scenario in this case as in figure \ref{a3q2t}, which evidently shows almost the same behaviour as in figure \ref{q1a3t} for the linear case. 
So we again find that the EOS parameter takes up values near -1 and stays on it for the eventuality, while the squared sound speed takes a few small jumps initially and then stays on values $\sim 0$. Hence, we again find that the logamediate models are suitable for all the holographic scenarios discussed. 
\\
\\
As we find the Logamediate scenario to also have the distinction of having both viable EOS parameter and squared speed sound values in various cases, we will also check the validity of the generalized second law $\Dot{S} \geq 0 $ for them. We find that the logamediate models, in all cases of Barrow, Tsallis and PLEC have a similar trend to each other.  
\begin{figure}[!ht]
    \centering
    \includegraphics[width=1\linewidth]{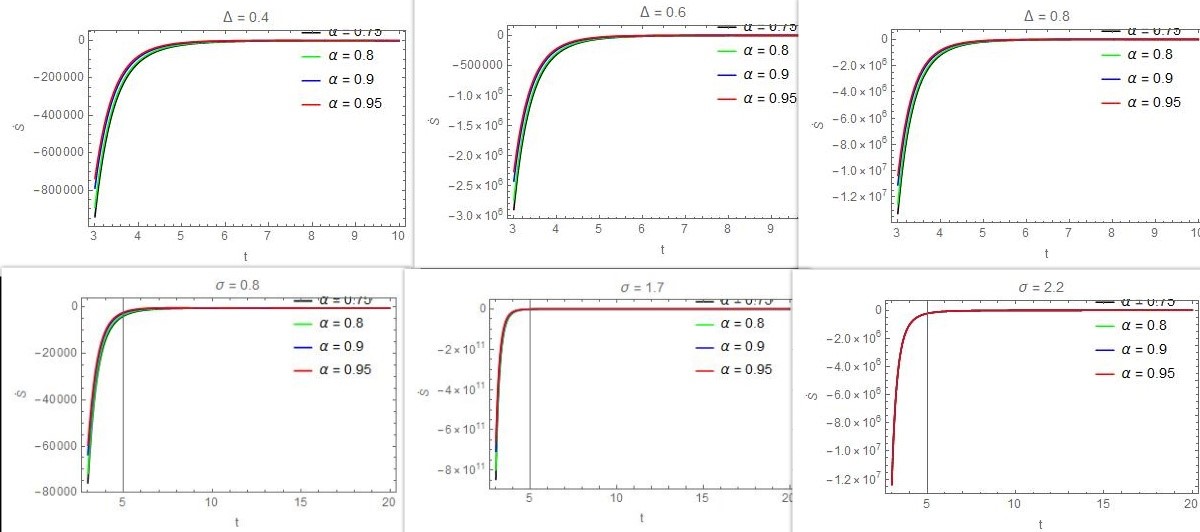}
    \caption{Plot of the time derivative of the universal entropy against time in Logamediate scenario for Tsallis and Barrow models}
    \label{sa3}
\end{figure}
In all cases, they start off from negative values but then eventually stay on for values $\sim 0$ and we show the plot for the same for Tsallis and Barrow scenarios in the linear interaction in figure \ref{sa3}. The trend is the same for the nonlinear interaction too and also the same for the PLEC models. So we can say that the logamediate models are the best fit seeing all this, as they produce viable EOS parameter and squared sound speed values in multiple models and are also  fit with generalized second law of thermodynamics.
\section{Conclusions}
In this work we have performed an extensive and detailed analysis of ansatz' based approach to holographic dark energy models. We considered three widely studied forms of extended HDE models, namely the Tsallis, Barrow and PLEC HDE models. We also considered 4 different forms of cosmological evolutions in our work which were namely the simple power law form, logamediate form, intermediate form and emergent universe form. We firstly developed the formulation for various key cosmological parametes in this case and also the key thermodynamical asepcts of these paradigms. 
\\
\\
The main takeaways from this are simple power law models perform horribly in all the HDE models, for both the linear and non-linear interaction regime as in all universal evolution scenarios, they do not give compatible models  Logamediate models perform the best Emergent scenario models do reasonably well too, a lot better than simple power law
Intermediate scenario models are better than simple power law but are still quite far behind the emergent scenario and logamediate models
While they produce subtle and minor difference, the results are very much equivalent for both linear and non-linear interaction schemes of DE and DM
The generalized second law of thermodynamics is satisfied in the viable HDE scenarios ( Logamediate models for all HDE scenarios and emergent universe models for certain Tsallis scenarios), but only just as the time derivative of entropy maintains values close to 0 in universal evolution making it just about satisfying the law. 
\\
\\
\section{Acknowledgements}
The research by M.K. was carried out at Southern Federal University with financial support from the Ministry of Science and Higher Education of the Russian Federation (State contract GZ0110/23-10-IF). The authors would like to thank Alexander Oliveros for various helpful discussions. 
\bibliography{Jreferences}
\bibliographystyle{unsrt}

\end{document}